\documentclass[osajnl,twocolumn,showpacs,10pt]{revtex4-1}
\usepackage{graphicx}
\usepackage{epsfig,bm}
\usepackage{amsfonts}
\usepackage{color}
\usepackage{amsmath}
\usepackage{url}
\usepackage{bbm}
\usepackage[normalem]{ulem}

\begin{document}
\title{Coherence effects in scattering order expansion of light by atomic clouds}

\author{Mohamed-Taha~\surname{Rouabah}$^{1,2}$, Marina~\surname{Samoylova}$^{3}$, Romain~\surname{Bachelard}$^{4}$,
Philippe W.~\surname{Courteille}$^{4}$,
Robin~\surname{Kaiser}$^{1}$,
Nicola~\surname{Piovella}$^{3}$}\email{Corresponding author:
nicola.piovella@unimi.it}
 \affiliation{
$^{1}$Universit\'{e} de Nice Sophia Antipolis, CNRS, Institut Non-Lin\'{e}aire de Nice, UMR 7335, F-06560 Valbonne, France\\
$^{2}$ Laboratoire de Physique Mathematique et Physique Subatomique, Universit\'{e} Constantine 1, Route Ain El Bey, 25017 Constantine, Algeria\\
$^{3}$ Dipartimento di Fisica, Universit\`{a} degli Studi di Milano, Via Celoria 16, I-20133 Milano, Italy\\
$^{4}$Instituto de F\'{i}sica de S\~{a}o Carlos, Universidade de
S\~{a}o Paulo,13560-970 S\~{a}o Carlos, SP, Brazil}

\begin{abstract}
We interpret cooperative scattering by a collection of cold atoms
as a multiple scattering process. Starting from microscopic
equations describing the response of $N$ atoms to a probe light
beam, we  represent the total scattered field as an infinite
series of multiple scattering events. As an application of the
method, we obtain analytical expressions of the coherent intensity
in the double scattering approximation for Gaussian density
profiles. In particular, we quantify the contributions of coherent
backward and forward scattering.
\end{abstract}


\maketitle

\section{Introduction}

Multiple scattering of light in disordered media has been
investigated since a long time using different
approaches~\cite{Rossum1999,Mishchenko2006,Akkermans2007}. Some of
them used coupled dipoles methods to describe light scattering by
dielectric particles, \cite{Purcell1973} while other approaches
interpret multiple scattering as a random walk of particle-like
photons where interference is neglected. This random walk is
described by a radiative transfer equation~\cite{Ishimaru1978,
Ishimaru1990,vandeHulst1990} which has been used for decades in
astrophysics, where the diffusive behavior is considered as a good
description of the light propagation. However, this approach must
be corrected when the scattered light wave emerges from the medium
in the backward direction. In this case, constructive
interferences arise and must be taken into account in order to
explain the enhancement of the backscattered intensity with
respect to the classical
prediction~\cite{Akkermans1986,Wolf1985,Albada1985,Lagendijk1988}.
This coherent back-scattering (CBS) has been observed for light
waves in a variety of
 media such as powder suspensions, biological
tissues or Saturn's rings as well as for laser-cooled atomic gases
\cite{Labeyrie1999,Bidel2002,Labeyrie2003}. The latter systems provide
an opportunity to observe cooperative effects in the light
scattering due to the absence of Doppler broadening.

Recently, a microscopic model of cooperative scattering by cold
atoms was
proposed~\cite{Morice1995,Svidzinsky2010,Courteille2010,Bienaime2011,Bachelard2011}
, which accounts for the interference effects. Signatures of
cooperativity have been observed in the reduction of the radiation
pressure force exerted on the center-of-mass of the atomic
cloud~\cite{Bienaime2010,Bender2010}. The microscopic model
provides an exact description of the scattering of a probe light
beam by $N$ atoms, i.e., taking into account interferences. It assumes the incident light beam to be weak
enough to neglect nonlinear effects, but naturally embeds the
multiple scattering process of the incoming photons bouncing among
the atoms, since the single-atom response is proportional to the
sum of the incident field and the field scattered by the other
atoms.

The aim of the present paper is to characterize the multiple
scattering nature of cooperative scattering, describing it not
from the point of view of the atoms, but of the scattered field.
Under this view, cooperative scattering appears as a sequence of
multiple scattering events where the emitted field is expressed as
the sum of successively scattered fields. This approach is of
particular importance for two reasons. Firstly, scattering in
optically dilute systems is very well described by a few
scattering events. The number of these events necessary to
reconstruct the solution is directly connected to the convergence
of the multiple scattering series. Secondly, it happens that the
multiple scattering expansion diverges in the optically dense
regime. This suggests that the interpretation of multiple
scattering as photons wandering from one atom to another one
starts to be incomplete. In such a regime, we can only talk about
global scattering by the entire cloud, and we lose track of the
light propagating in the cloud at different orders of scattering.
As a consequence, the scattered field seen by each atom can not be
obtained as the coherent sum over all the light trajectories, but
it must result from a global approach, determining  or the
single-atom response to the total scattered field
\cite{Bienaime2013} or the eigenmodes of the system
\cite{Sokolov2009}.

Let us outline that we treat the light scattering  \textit{ab
initio}, i.e., considering point-like atoms in the vacuum. This is
different from the common approach resorting to an effective
Green's function, where the average atomic medium is described by
a refractive index, which implies the introduction of a mean free
path \cite{Ishimaru1978,Lagendijk1988,Nieuwenhuizen1992}. On the
contrary, in our model the refractive index emerges \textit{a
posteriori} as a result of the multiple light scattering
process~\cite{Bachelard2012}.

The paper is organized as follows. In Sec.~\ref{sec:micromodel} we
review the cooperative scattering model using the more general
vectorial model, and show how the atomic response builds up as a
reaction to both the incident and scattered fields. The multiple
scattering approach is presented in Sec.~\ref{sec:MS}. In
Sec.~\ref{sec:CBS} we discuss CBS in the double scattering
approximation, deriving analytical expressions for a gaussian
sphere. We also demonstrate how coherent multiple scattering
theory allows to obtain corrections to the single-scattering
forward emission.

\section{Microscopic approach to the cooperative scattering\label{sec:micromodel}}

The cooperative scattering by $N$ atoms with fixed positions
$\mathbf{r}_j$ and illuminated by a monochromatic light beam with
electric field components $E_{in}^\alpha(\mathbf{r})\exp(-i\omega
t)$ is described by the following set of coupled
equations~\cite{Courteille2010,Svidzinsky2010}:
\begin{eqnarray}\label{betajk}
    \frac{db_j^{\alpha}}{dt}&=&\left(i\Delta-\gamma/2\right)b_j^{\alpha}
   -i(d/\hbar)E_{in}^{\alpha}(\mathbf{r}_j)\nonumber\\
    &-&(\gamma/2)\sum_{\alpha'}\sum_{m\neq
j}G_{\alpha,\alpha'}(\mathbf{r}_{j}-\mathbf{r}_{m})b_m^{\alpha'},
\end{eqnarray}
where $d$ is electric dipole matrix element,
$\Delta=\omega-\omega_a$ is the detuning of the incident light
frequency $\omega=ck$ from the atomic resonance frequency
$\omega_a$ and $\gamma=d^2k^3/3\pi\epsilon_0\hbar$ is the
spontaneous decay rate. In the right-hand side of
Eq.~\eqref{betajk} the first term describes the single-atom
dynamics, the  second term corresponds to the external field and
the last term describes the radiation of all other atoms on
$j$th-atom. Eq.~\eqref{betajk} is derived from a quantum approach
modelling the scattering of a single photon as being scattered in
a mode tailored by the spatial atomic
distribution~\cite{Friedberg1973,Morice1995,Manassah2012b}, but
also from a classical approach where the atoms are considered as
oscillating dipoles induced by a classical laser field described
by Maxwell equations~\citep{Svidzinsky2010}. The $j$th atom
experiences electric dipole transitions between the single ground
state $|g_j\rangle$ and the degenerate triplet excited state
$|e_j^{\alpha}\rangle$, where $\alpha=x,y,z$ and $b_j^\alpha$ are
the probability amplitudes of the single-excitation atomic state
$|\Psi\rangle_{\mathrm{e}}=\exp(-i\Delta t)\sum_j\sum_{\alpha}
b_j^\alpha|g_1,\dots,e_j^\alpha,\dots,g_N\rangle$.
$G_{\alpha,\alpha'}$ are the components of the symmetric tensor:
\begin{eqnarray}\label{Gjm}
    G_{\alpha,\alpha'}(\mathbf{r}) &=& \frac{3}{2}\frac{e^{ikr}}{ikr}\left\{\left[\delta_{\alpha,\alpha'}-\hat n_\alpha\hat n_{\alpha'}\right]\right.
    \nonumber\\
    &+&\left.\left[\delta_{\alpha,\alpha'}-3\hat n_\alpha\hat n_{\alpha'}\right]\left[i/(kr)-1/(kr)^2\right]\right\}\nonumber\\
\end{eqnarray}
with $r=|\mathbf{r}|$ and $\hat n_\alpha$ being the components of
the unit vector $\mathbf{\hat n}=\mathbf{r}/r$. The vectorial
Green's function (\ref{Gjm}) can be obtained from the scalar
Green's function $G(r)=\exp(ikr)/(ikr)$:
\begin{equation}\label{GvGs}
    G_{\alpha,\alpha'}(\mathbf{r})=\frac{3}{2}\left[\delta_{\alpha,\alpha'}+\frac{1}{k^2}\frac{\partial^2}{\partial\alpha\partial {\alpha'}}\right]G(r).
\end{equation}
The steady-state problem of Eq.(\ref{betajk}) boils down to a
linear one for the complex vectors $\mathbf{b}_j$ with spatial
components $b_j^{\alpha}$ :
\begin{equation}\label{betasta}
   \mathbf{b}_j=\frac{1}{\Delta+i\gamma/2}\left[\frac{d}{\hbar}\mathbf{E}_{\mathrm{in}}(\mathbf{r}_j)-i\frac{\gamma}{2}
    \sum_{m\neq j}\mathbf{G}(\mathbf{r}_{j}-\mathbf{r}_m)\cdot
    \mathbf{b}_m\right].
\end{equation}
Giving the atomic positions $\mathbf{r}_j$ and incident field
$\mathbf{E}_{\mathrm{in}}(\mathbf{r}_j)$, it can be solved
numerically by inverting a $3N\times 3N$ symmetric matrix.

The scattered field at a position $\mathbf{r}$ is derived from $\mathbf{b}_j$ using the microscopic Maxwell equations for sources of polarization
$\mathbf{P}(\mathbf{r})=-d\sum_j\mathbf{b}_j\delta(\mathbf{r}-\mathbf{r}_j)$.
The result, as demonstrated in the Appendix \ref{Appendix:field},
reads:
\begin{equation}\label{Evetto}
    \mathbf{E}_\mathrm{sca}(\mathbf{r})=-i\frac{dk^3}{6\pi\epsilon_0}\sum_{m=1}^N\mathbf{G}(\mathbf{r}-\mathbf{r}_m)\cdot\mathbf{b}_m.
\end{equation}
The scattered field $\mathbf{E}_\mathrm{sca}(\mathbf{r}_j)$ at the atomic
position $\mathbf{r}_j$ has a divergent contribution in the term
$m=j$ of the sum in Eq.\eqref{Evetto}. This is a well-known
problem of the self-field, i.e., the field generated by the atom
acting on the atom itself. Usually, in multiple scattering
theories this problem is circumvented by introducing a cut-off
length of the order of the size of the real physical scatterer
\cite{Nieuwenhuizen1992}. However, in the present approach the
self-field does not play any role. In fact, calling
$\mathbf{E}_{\mathrm{self}}(\mathbf{r}_j)$ the self-field of the
atom $j$,  Eq.\eqref{Evetto} turns into:
\begin{equation}\label{Esca}
    \mathbf{E}_\mathrm{sca}(\mathbf{r}_j)=\mathbf{E}_{\mathrm{self}}(\mathbf{r}_j)-i\frac{dk^3}
    {6\pi\epsilon_0}\sum_{m\neq j}\mathbf{G}(\mathbf{r}_j-\mathbf{r}_m)\cdot\mathbf{b}_m.
\end{equation}
Combining Eqs.(\ref{betasta}) and (\ref{Esca}), one can obtain:
\begin{equation}\label{beta+E}
   \mathbf{b}_j=\frac{d}{\hbar\left(\Delta+i\gamma/2\right)}
   \left[\mathbf{E}_{\mathrm{in}}(\mathbf{r}_j)+\mathbf{\bar E}(\mathbf{r}_j)\right],
\end{equation}
where $\mathbf{\bar
E}(\mathbf{r}_j)=\mathbf{E}_\mathrm{sca}(\mathbf{r}_j)-\mathbf{E}_{\mathrm{self}}(\mathbf{r}_j)$
is the electric field acting on the $j$th atom without the
self-field contribution. $\mathbf{\bar E}(\mathbf{r}_j)$ is
introduced to describe the field at the atomic positions and avoid
the divergence problem present in \eqref{Evetto}. The electric
dipole moment of each atom $\mathbf{p}_j=-d\mathbf{b}_j$ is
directly proportional to the sum of the incident field and the one
scattered by all  other atoms, as assumed in the cooperative
scattering description~\eqref{betasta}.

\section{The multiple scattering series\label{sec:MS}}

In the microscopic approach of cooperative scattering presented in
Sec.~\ref{sec:micromodel}, the radiation field is determined from
the knowledge of the individual atomic responses $\mathbf{b}_j$,
which are themselves derived from the linear problem
Eq.\eqref{betasta}. On the contrary, the multiple scattering
approach is based on a recursive set of equations for the sole
radiation field. It is obtained by inserting Eq.\eqref{beta+E}
back into Eq.\eqref{Esca}, leading to an implicit equation for the
scattered field $\mathbf{\bar E}(\mathbf{r}_j)$ acting on the
$j$th atom:
\begin{equation}\label{Evectot}
    \mathbf{\bar E}(\mathbf{r}_j)=
    \kappa(\delta)\sum_{m\neq j}\mathbf{G}(\mathbf{r}_j-\mathbf{\mathbf{r}}_m)\cdot\left[\mathbf{E}_{\mathrm{in}}
    (\mathbf{r}_m)+\mathbf{\bar E}(\mathbf{r}_m)\right],
\end{equation}
where $\kappa(\delta)=1/(2i\delta-1)$ and $\delta=\Delta/\gamma$. Introducing the total field $\mathbf{ \bar
E}_{tot}(\mathbf{r}_j)=\mathbf{E}_{in}(\mathbf{r}_j)+\mathbf{\bar
E}(\mathbf{r}_j)$, the above equation can also be written in a matrix form
\begin{equation}\label{Etot:matrix}
\mathbf{\bar E}_{\mathrm{tot}} = (\mathbf{I} - \mathcal{G})^{-1}
~\mathbf{E}_{\mathrm{in}},
\end{equation}
where $\mathbf{\bar E}_{\mathrm{tot}}$ and $\mathbf{ E}_{in}$ are
vectors containing the $3N$ components of the effective electric
field acting on the $j$th atom (without the self-field
contribution) and the incident field, respectively; $I$ is the $3N
\times 3N$ unit matrix, and $\mathcal{G} =
\kappa(\delta)\mathbf{G}$ is a $3N \times 3N$ matrix containing
the Green's function $\mathbf{G}_{jm} = \mathbf{G}(\mathbf{r}_j -
\mathbf{r}_m)$ whose component are given by Eq.\eqref{Gjm}.

The multiple scattering approach consists in solving
Eq.(\ref{Evectot}) by iteration. Introducing the scattered field
$\mathbf{\bar E}^{(n)}$ after $n$ scattering events, the following
recurrence relation is obtained from Eq.~\eqref{Etot:matrix}:
\begin{equation}\label{En}
    \mathbf{\bar E}^{(n)}(\mathbf{r}_j)=\kappa(\delta)\sum_{m\neq j}
    \mathbf{G}(\mathbf{r}_j-\mathbf{\mathbf{r}}_{m})\cdot\mathbf{\bar E}^{(n-1)}(\mathbf{r}_m),
\end{equation}
where $n=1,2\dots$ and the incident field $\mathbf{\bar
E}^{(0)}(\mathbf{r}_j)=\mathbf{E}_{in}(\mathbf{r}_j)$ plays the role
of the seed. The total scattered field \eqref{Evectot} corresponds to the infinite
sum of all scattered fields
\begin{equation}\label{Eseries}
    \mathbf{\bar E}(\mathbf{r}_j)=\sum_{n=1}^\infty
    \mathbf{\bar E}^{(n)}(\mathbf{r}_j),
\end{equation}
provided the series is converging. The effective field felt by the scatterer at $\mathbf{r}_j$ consists of the incident wave
$\mathbf{E}_{in}(\mathbf{r}_j)$ and the wave scattered from the
other atoms in the cloud (except the self field of the atom at
$\mathbf{r}_j$) given by Eq.\eqref{Eseries} and resulting from an
increasing number of scattering events of the incident field. Eq.~\eqref{Etot:matrix} can be extended as a series:
\begin{eqnarray}\label{Etot:Exp}
    \mathbf{\bar E}_{\mathrm{tot}}(\mathbf{r}_j)&=&\mathbf{E}_{\mathrm{in}}(\mathbf{r}_j)+
    \kappa(\delta)\sum_{m\neq j} \mathbf{G}(\mathbf{r}_j-\mathbf{r}_m)\mathbf{E}_{\mathrm{in}}(\mathbf{r}_m) \nonumber\\
    &+&  \kappa^2(\delta)\sum_{m\neq j} \mathbf{G}(\mathbf{r}_j-\mathbf{r}_m)\sum_{l\neq m}
    \mathbf{G}(\mathbf{r}_m-\mathbf{r}_l)\mathbf{E}_{\mathrm{in}}(\mathbf{r}_l)  \nonumber\\
    &+& \dots 
\end{eqnarray}
The infinite series \eqref{Eseries} converges only if all eigenvalues
of $\mathcal{G}$ have their modulus less than unity
\cite{Fadeeva1959}. When this condition is satisfied, the multiple
scattering expansion can be used to calculate the radiated field.

\begin{figure}
\centering\includegraphics[width=0.98\linewidth]{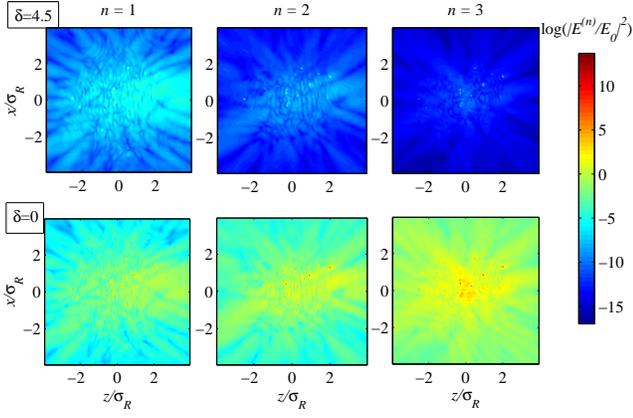}
\caption{\label{convdivfield}(Color online) Profile of the
radiation field inside a Gaussian cloud,
$\mathbf{E}^{(n)}(\mathbf{r}) = \kappa(\delta)
\sum_{j}\mathbf{G}(\mathbf{r}-\mathbf{r}_j)
\mathbf{\bar{E}}^{(n-1)}(\mathbf{r}_j)$ in the $y=0$ plane for
different orders $n$ (from left to right) and two different
optical thicknesses (top and bottom). For small optical thickness
($b(\delta) = 0.061$, top line), the field decreases as the
scattering order $n$ increases, and the series~\eqref{Eseries}
converges. For larger optical thickness ($b(\delta)=5$, bottom
line), the presence of eigenvalues of modulus larger than unity
makes the multiple scattering series to diverge. In both cases,
the presence of local fields much stronger than the incident one
is due to the divergent field radiated in the vicinity of the
atoms, which can be arbitrary close to the $y=0$ plane.
Simulations realized for a Gaussian cloud of $N=500$ atoms with an
on-resonant optical thickness $b_0=5$ and standard deviation
$\sigma_R= 17.32/k$, where $b_0=3N/(k\sigma_R)^2$; top pictures
correspond to $\delta=4.5$ and $b(\delta) = 0.061$, bottom
pictures to the resonant case $\delta=0$ and
$b(\delta)=5$.}\vspace{-0.4cm}
\end{figure}

\begin{figure}
\centering\includegraphics[width=0.98\linewidth]{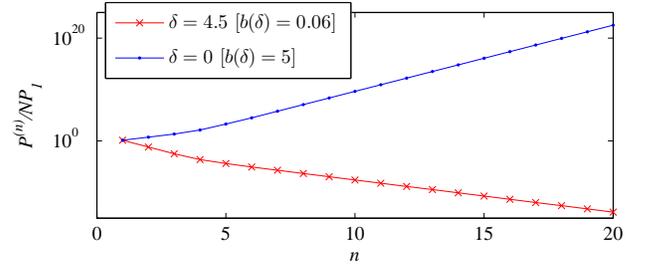}
\caption{\label{f:convdivfield}(Color online) Far-field power
$P^{(n)}=(\varepsilon_0 c/2)\int |\mathbf{E}^{(n)}|^2\mbox{d}S$
radiated by the atomic cloud (measured on a spherical surface of
radius $r \gg k$ ) vs. the scattering order $n$  for the same
parameters as in Fig.\ref{convdivfield}. For small optical
thickness ($b(\delta) = 0.061$, red crosses), the scattered power
decreases as the scattering order $n$ increases, whereas for
larger optical thickness ($b(\delta)=5$, blue dots), the power
diverges. $P^{(n)}$ is in units of the independent-atom power
$NP_1$, where $P_1=(4\pi I_0/k^2)/(1+4\delta^2)$ and $I_0$ is the
incident intensity.}\vspace{-0.4cm}
\end{figure}

Note that the convergence of the series \eqref{Eseries} or of the
sum \eqref{Etot:Exp} is not tied to the existence of a solution
for the field. Indeed Eq.\eqref{Evectot} always admits a solution,
whereas the linear operator $\mathcal{G}$ of the recurrence
Eq.\eqref{En} may admit eigenvalues of modulus larger than unity,
in which case Eq.\eqref{Eseries} does not converge. In that case
each scattering order radiates more light than the previous one,
and the multiple scattering expansion diverges. In order to
illustrate this point, the electric field profile inside a
Gaussian cloud $\mathbf{E}^{(n)}(\mathbf{r}) = \kappa(\delta)
\sum_{j}\mathbf{G}(\mathbf{r}-\mathbf{r}_j)
\mathbf{\bar{E}}^{(n-1)}(\mathbf{r}_j)$ and the far-field radiated
power $P^{(n)}=(\varepsilon_0 c/2)\int
|\mathbf{E}^{(n)}|^2\mbox{d}S$ are plotted for different orders
$n$ in Fig. \ref{convdivfield} and \ref{f:convdivfield}
respectively. Both quantities have been obtained for two different
optical thicknesses $b(\delta)=b_0/(1+4\delta^2)$, where
$b_0=3N/(k\sigma_R)^2$ is the on-resonant optical thickness for a
Gaussian cloud with rms size $\sigma_R$. The two simulations have
been realized for $N=500$ atoms with $b_0=5$, $\sigma_R= 17.32/k$
and for two different detuning, $\delta=4.5$ and $\delta=0$,
corresponding to $b(\delta) = 0.061$ and $b(\delta)=5$,
respectively. For the case of small optical thickness ($b(\delta)
= 0.061$), the field decreases as the scattering order $n$
increases, and the series \eqref{Eseries} converges. For the case
of larger optical thickness ($b(\delta)=5$), the presence of
eigenvalues of modulus larger than unity makes the multiple
scattering series diverge. Hence, in presence of above-unity
eigenvalues of $\mathcal{G}$, the multiple scattering description
loses its validity: for sufficiently dense media, due to the
long-range interaction of the Green's function, the build-up of
the scattered radiation field cannot be seen as the sum of
interactions involving an increasing number of atoms, and the
local iteration of the scattering event described by Eq.\eqref{En}
is no longer possible. Instead, the total scattered field is a
result of a global interaction with the entire sample. Let us
remark that the criterion of all eigenvalues having modulus below
unity for the convergence of the series is in agreement with the
results of Ref.\cite{Bohren1988}. A detailed study of the typical
spectrum of the linear operator in \eqref{En} has been proposed in
\cite{Skipetrov2011b}, yet it is important to mention that the
spectrum exhibits strong fluctuations from one realization to
another. Since the multiple scattering process corresponds to a
geometric series, the radiated power grows or decreases as a
power-law of the largest eigenvalue of the linear operator in
Eq.\eqref{En} for large $n$.

Once obtained from Eq.(\ref{Eseries}) $\mathbf{\bar
E}(\mathbf{r}_j)$, using Eqs.(\ref{Evetto}) and (\ref{beta+E}) the
scattered field in position $\mathbf{r}\neq\mathbf{r}_j$ is fully
determined as:
\begin{eqnarray}\label{Exter}
    \mathbf{E}_{\mathrm{sca}}(\mathbf{r})&=&
    \kappa(\delta)\sum_{j=1}^N\mathbf{G}(\mathbf{r}-\mathbf{\mathbf{r}}_j)
    \cdot\left[\mathbf{E}_{\mathrm{in}}(\mathbf{r}_j)+\mathbf{\bar E}(\mathbf{r}_j)\right].
\end{eqnarray}
The multiple scattering nature of the field detected at
$\mathbf{r}$ is evident from Eq.(\ref{Exter}): the first term in the sum
represents all the single scattering and the second term collects
all the multiple scattering. The $\mathbf{\bar E}(\mathbf{r}_j)$ term, which contains all
scattering orders starting from the first (see Eq.~\eqref{Eseries}), yields the double and higher scattering
orders in \eqref{Exter} after applying $G$.

We point out that the solution for the scattered field (given by
the infinite series of Eq.(\ref{Eseries}) and Eq.(\ref{Exter})) is
fully equivalent to solving Eq.\eqref{betasta} for $\mathbf{b}_j$
and then calculating the field using Eq.(\ref{Evetto}).
Eq.~\eqref{betasta} can be solved exactly by numerical inversion
of this linear problem. The only constraint we deal with is
 the limited number $N$ of scatterers that
can be handled by the computer capacities. From the perspective of
computing the scattered field, the microscopic approach has a
clear advantage over the multiple scattering one, which requires
the evaluation of an infinite sum. The numerical solution of the
microscopic approach provides a solution valid for arbitrary
distributions of scatterers in the vacuum using only finite
matrices. Finally, it treats light as a complex field, not only as
an intensity, so that it naturally embeds the coherence of the
multiple scattering process.

In the far-field limit, the scattered field can be derived using
the asymptotic form of the vectorial Green's function for $r\gg
r_j$:
\begin{equation}\label{Gfar}
    G_{\alpha,\alpha'}(\mathbf{r}-\mathbf{r}_j)\approx\ \frac{3}{2}\frac{e^{ikr}}{ikr} \left[\delta_{\alpha,\alpha'}-\hat n_\alpha\hat
    n_{\alpha'}\right]e^{-i\mathbf{k}\cdot \mathbf{r}_j},
\end{equation}
where $\mathbf{k}=k\mathbf{\hat n}$. For an incident plane wave
with $\mathbf{E}_{in}(\mathbf{r})=\mathbf{\hat
e_0}E_0\exp(i\mathbf{k}_0\cdot \mathbf{r})$, where $\mathbf{\hat
e}_0$ is the unit polarization vector, the scattered far field derived
from Eq.(\ref{Exter}) is:
\begin{eqnarray}\label{Efar}
    \mathbf{E}_{\mathrm{sca}}^{\mathrm{far}}(\mathbf{r})&=&
    \mathbf{E}_{1}^{\mathrm{far}}(\mathbf{r})+\mathbf{E}_{\mathrm{ms}}^{\mathrm{far}}(\mathbf{r}),
\end{eqnarray}
where
\begin{equation}\label{E1far}
    \mathbf{E}_{1}^{\mathrm{far}}(\mathbf{r})=\kappa(\delta)E_0\frac{e^{ikr}}{ikr}
    [\mathbf{\hat n}\times(\mathbf{\hat n}\times\mathbf{\hat e}_{0})]NS_N(\mathbf{k}-\mathbf{k}_0),
\end{equation}
corresponds to the single scattering order and
$S_N(\mathbf{k}-\mathbf{k}_0)=(1/N)\sum_j\exp[-i(\mathbf{k}-\mathbf{k}_0)\cdot\mathbf{r}_j]$
is the structure factor. Eq.\eqref{E1far} is the well-known
expression for the Rayleigh scattering by particles with size much
smaller than the optical wavelength when each atom is excited by
the incident field only. Then the scattered field results from a
coherent superposition of the field amplitudes generated by each
atom and is proportional to the structure factor. The multiple
scattering contribution to the far field of Eq.(\ref{Efar}) is,
using Eq.(\ref{Eseries}),
\begin{equation}\label{Emsfar}
    \mathbf{E}_{\mathrm{ms}}^{\mathrm{far}}(\mathbf{r})=\kappa(\delta)\frac{e^{ikr}}{ikr}N\,
    \mathbf{\hat n}\times[\mathbf{\hat n}\times\mathbf{F}(\mathbf{k})],
\end{equation}
\vspace{-0.3cm}
where
\begin{equation}\label{Fk}
    \mathbf{F}(\mathbf{k})=\frac{1}{N}\sum_{j=1}^N
    \mathbf{\bar E}(\mathbf{r}_j)e^{-i\mathbf{k}\cdot
    \mathbf{r}_j}.
\end{equation}
We stress that our approach is valid beyond the single-scattering
limit. The single-scattering approximation holds when the optical
thickness $b(\delta)=b_0/(1+4\delta^2)$ is much smaller than
unity. In contrast, our multiple scattering approach is valid for
finite values of $b(\delta)<1$:  the convergence of the series of
Eq.\eqref{Eseries} guarantees the validity of our multiple
scattering expansion \cite{Akkermans2008,Skipetrov2011}.

\begin{figure}\centering
\includegraphics[width=1\linewidth]{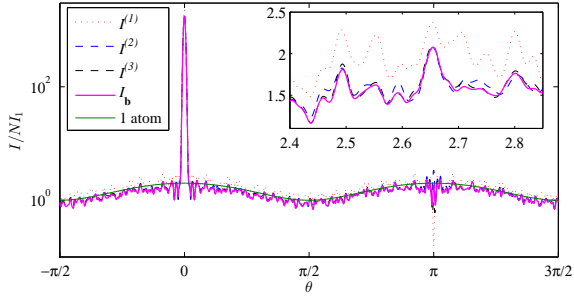}
\caption{\label{f:ConvEbeta}(Color online) Intensity diagram,
$I^{(n)}=(\varepsilon_0c/2)|\sum_{j=1}^n \mathbf{E}^{(j)}|^2$ vs
the polar angle $\theta$, for $n=1,2,3$, derived from
Eq.\eqref{Exter}. The intensity scattered by the dipoles,
$I_\mathbf{b}=(\varepsilon_0c/2)|\mathbf{E}_\mathbf{b}|^2$, is
derived from Eq.~\eqref{betasta} and contains all the scattering
orders. The modulation of the background is due to the vectorial
nature of the light (linearly polarized light): the single-atom
intensity $I_1(1+\cos^2\theta)$  is plotted as a plain green line,
where $I_1 = I_0/[k^2r^2)(1+4\delta^2)]$. The inset shows a zoom
of the radiation (linear scale). Simulations have been
realized for a Gaussian cloud of $N=1000$ atoms with on-resonance
optical thickness $b_0=1$, detuning $\delta=0$ and rms size
$\sigma_R \approx 54.8/k$ given by $b_0 = 3N/(k\sigma_R)^2$. The
intensity is averaged over the azimuthal angle $\phi$ and is in
unit of $NI_1$.}\vspace{-0.4cm}
\end{figure}

Finally, we emphasize that if the infinite sum in
Eq.~\eqref{Eseries} converges, it gives the {\it exact} solution
for the collective scattering problem given by
Eqs.~(\ref{betasta}--\ref{Evetto}). To illustrate this point, we
compare the intensity radiated up to the $n$th order
$I^{(n)}=(\varepsilon_0 c/2)|\sum_{j=1}^n \mathbf{E}^{(j)}|^2$ for
$n=1,2,3$, as well as the intensity $I_\mathbf{b}$ provided by the dipole
amplitudes $\mathbf{b}_m$ derived from Eq.\eqref{betasta} and
containing all the scattering orders. In particular, $I^{(1)}$ describes the single
scattering only, $I^{(2)}$ the sum of single and double scattering, etc. For an optical thickness
$b(\delta)$  equal to unity ($b_0=1$, $\delta=0$), the convergence
is relatively slow, but clearly visible in Fig.~\ref{f:ConvEbeta}.
For optical thickness much smaller than unity, the convergence is
very fast and the single-scattering physics contained in
$\mathbf{E}^{(1)}$ describes already very well the total scattered
field (not shown here).

We note in Fig.~\ref{f:ConvEbeta} that the coupled dipole equation predicts a background radiation
lower than $N$ times that of a single-atom. This suggests a reduction of the background radiation,
in favor of the coherent forward radiation, under the effects of cooperativity. This effect will be
the subject of a future dedicated study.

\section{Coherent Backward and Forward Scattering\label{sec:CBS}}

As an application of the multiple scattering approach, we
investigate coherent backscattering (CBS) and coherent forward scattering (CFS) from a collection of $N$ atoms.
For the sake of simplicity, we assume that the radiation
waves are scalar, neglecting polarization and near-field effects.
In the scalar radiation theory,  the three components $b_j^\alpha$
in Eq.~\eqref{betajk} are replaced by a single value $\beta_j$,
the vectorial kernel $G_{\alpha,\alpha'}(\mathbf{r})$ is
substituted by the scalar Green's function $G(r)=\exp(ikr)/(ikr)$
and the decay constant $\gamma$ is replaced by
$\Gamma=(3/2)\gamma$
\cite{Bienaime2011,Scully2006,Svidzinsky2008}. Then, the scalar
equivalent of Eq.~\eqref{Exter} is
\begin{eqnarray}\label{Extra:scalar}
    E_{\mathrm{sca}}(\mathbf{r})&=&\kappa(\delta)\sum_{j=1}^NG(|\mathbf{r}-\mathbf{\mathbf{r}}_j|)
    \left[E_{\mathrm{in}}(\mathbf{r}_j)+\bar E(\mathbf{r}_j)\right].\nonumber\\
\end{eqnarray}
We approximate the multiple scattering field $\bar
E(\mathbf{r}_j)$ by its first contribution:
\begin{equation}\label{E1}
    \bar E(\mathbf{r}_j)\approx\kappa(\delta)\sum_{m\neq j}
    G(|\mathbf{r}_j-\mathbf{\mathbf{r}}_m|)E_{\mathrm{in}}(\mathbf{r}_m),
\end{equation}
which is equivalent to considering single- and double-scattering
events only:
\begin{eqnarray}\label{Etot}
    E_{\mathrm{tot}}(\mathbf{r})&=&E_{\mathrm{in}}(\mathbf{r})
    +\kappa(\delta)\sum_{j=1}^NG(|\mathbf{r}-\mathbf{\mathbf{r}}_j|)E_{\mathrm{in}}(\mathbf{r}_j)\nonumber\\
    &+& \kappa^2(\delta)\sum_{m=1}^N \sum_{j\neq m}G(|\mathbf{r}-\mathbf{\mathbf{r}}_m|)G(|\mathbf{r}_m-\mathbf{\mathbf{r}}_j|)
    E_{\mathrm{in}}(\mathbf{r}_j).\nonumber\\
\end{eqnarray}

\begin{figure}
    \centering\includegraphics[width=0.80\linewidth]{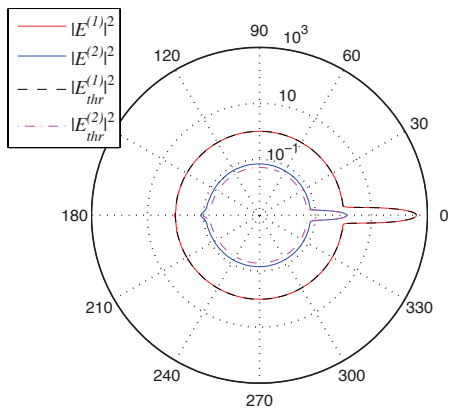}
    \caption{\label{f:DiagEm3o}(Color online) Radiation profile $|E^{(n)}(\theta)|^2$ in the far-field limit for scattering orders $n=1,\ 2$.
    The single-scattering order $E^{(1)}$ exhibits only a forward contribution (peaks pointing to the right) and a homogeneous background.
    The double-scattering contribution $E^{(2)}$ shows both forward and backward (CBS) patterns (peaks pointing to the left)
    in addition to the background. The theoretical curves ($thr$) are derived from Eqs.~\eqref{Tfin}, \eqref{SGauss} and \eqref{TGauss}.
    Simulations are realized for a Gaussian spherical cloud consisting of $N=400$ atoms with resonant optical thickness
    $b_0=2N/(k\sigma_R)^2 = 1$, standard deviation $\sigma_R \approx 28.3/k$, detuning $\delta=1$ and averaged over $1000$ realizations.
    The incoming field is unity $E_0=1$ and the radius of observation is $3\cdot10^4/k$. Scale is Logarithmic.}\vspace{-0.5cm}
\end{figure}

The second term in Eq.~\eqref{Etot} describes the single
scattering of the incident wave by each atom  in position
$\mathbf{r}_j$, followed by its propagation towards $\mathbf{r}$.
The third term corresponds to the double scattering contribution,
i.e. the photons are first scattered by the atoms in
$\mathbf{r}_j$, then propagate to $\mathbf{r}_m$, where they are
scattered again and reach position $\mathbf{r}$. As it can be
observed in Fig.~\ref{f:DiagEm3o}, the double-scattering is the
first of the multiple scattering process that contributes to CBS.
It results from the interference between the wave which is first
scattered in $\mathbf{r}_j$ and then in $\mathbf{r}_m$, and the
reciprocal path, when the wave is first scattered in
$\mathbf{r}_m$ and then in $\mathbf{r}_j$. This effect can be
captured by calculating the scattered field in the far-field
limit, approximating the Green's function as
$G(|\mathbf{r}-\mathbf{\mathbf{r}}_j|)\approx\exp(ikr-i\mathbf{k}\cdot
\mathbf{r}_j)/(ikr)$:
\begin{eqnarray}\label{Etot:far}
    E_{\mathrm{sca}}(\mathbf{r})&=&\kappa(\delta)\frac{e^{ikr}}{ikr}E_0\sum_{j=1}^Ne^{i(\mathbf{k}_0-\mathbf{k})\cdot\mathbf{r}_j}\nonumber\\
    &+& \kappa^2(\delta)\frac{e^{ikr}}{ikr}E_0\sum_{m=1}^N \sum_{j\neq m} G(|\mathbf{r}_m-\mathbf{\mathbf{r}}_j|)
    e^{i(\mathbf{k}_0\cdot\mathbf{r}_j-\mathbf{k}\cdot\mathbf{r}_m)},\nonumber\\
\end{eqnarray}
where we assume that the incident field is a plane--wave
$E_{\mathrm{in}}(\mathbf{r})=E_0\exp(i\mathbf{k}_0\cdot\mathbf{r})$.
Introducing the factor
\begin{eqnarray}\label{TN}
    T_N(\mathbf{k},\mathbf{k}_0)&=&\frac{1}{N}\sum_m\sum_{j\neq
    m}G(|\mathbf{r}_j-\mathbf{r}_m|)
    e^{i(\mathbf{k}_0\cdot\mathbf{r}_j-\mathbf{k}\cdot\mathbf{r}_m)},\nonumber\\
\end{eqnarray}
the scattered intensity up to the second scattering order can be
written as:
\begin{eqnarray}\label{Etot:far2}
    I_{\mathrm{sca}}(\mathbf{r})&=&I_1 N^2\left|S_N(\mathbf{k}-\mathbf{k}_0)+ \kappa(\delta)T_N(\mathbf{k},\mathbf{k}_0)\right|^2,
\end{eqnarray}
where $I_1=I_0/[k^2r^2(1+4\delta^2)]$ is the single-atom scattered
intensity and $I_0$ is the intensity of the incident wave. Upon configuration averaging, the structure factor
gives an incoherent contribution,
$|S_N(\mathbf{k}-\mathbf{k}_0)|^2=1/N$, while the coherent
contribution in the forward direction for large $N$ can be written
as a continuous integral:
\begin{equation}\label{SInfty}
    S_\infty=\frac{1}{N}\int d\mathbf{r}\rho(\mathbf{r})\exp[i(\mathbf{k}-\mathbf{k}_0)\cdot\mathbf{\mathbf{r}}],
\end{equation}
where $\rho(\mathbf{r})$ is the atomic density. $|T_N|^2$ yields
an incoherent contribution plus two coherent contributions which,
however, have different origins. Taking the square modulus of
Eq.\eqref{TN} and considering only equal pairs of atoms $(j,m)$ in
$T_N$ and $T_N^*$, we obtain:
\begin{equation}\label{TN2}
    |T_N(\mathbf{k},\mathbf{k}_0)|_{\mathrm{pair}}^2\approx\frac{1}{N^2}\sum_m\sum_{j\neq m}
        \frac{1+\cos[(\mathbf{k}+\mathbf{k}_0)\cdot(\mathbf{r}_j-\mathbf{r}_m)]}{k^2|\mathbf{r}_j-\mathbf{r}_m|^2}.
\end{equation}

The first incoherent term emerges when the same pair of atoms is
considered twice ($1/k^2|\mathbf{r}_j-\mathbf{r}_m|^2$ term in
Eq.\eqref{TN2}), whereas the second term results from the pair
$(j,m)$ crossed with its reciprocal path $(m,j)$ (cosine term).
The latter is known as the CBS term~\cite{Akkermans1986}, since it
also yields a backward coherent radiation. In the diagrammatic
approach, the first term in Eq.~\eqref{TN2} corresponds to the
first 'ladder' term and the second one to the 'most-crossed
term'~\cite{Nieuwenhuizen1992}. Besides these pair terms, $T_N$
also gives a coherent contribution to the forward intensity due to
the processes involving more than two atoms. This contribution in
the continuous density limit can be written as:
\begin{eqnarray}\label{Tcont}
    T_\infty(\mathbf{k},\mathbf{k}_0)&=&\frac{1}{N}\int d\mathbf{r}_1\rho(\mathbf{r}_1)\int d\mathbf{r}_2\rho(\mathbf{r}_2)
    \frac{\exp(ik|\mathbf{r}_1-\mathbf{r}_2|)}{ik|\mathbf{r}_1-\mathbf{r}_2|}\nonumber\\
    &\times & e^{i(\mathbf{k}_0\cdot\mathbf{r}_1-\mathbf{k}\cdot\mathbf{r}_2)}.\nonumber\\
\end{eqnarray}
Collecting the different contributions, the scattered intensity up
to the double scattering order reads:
\begin{eqnarray}\label{Etot:terms}
    I_{\mathrm{sca}}&=&
    I_1N\left\{1+\frac{N}{1+4\delta^2}|T_N|^2_{\mathrm{pair}}+N|F|^2\right\},
\end{eqnarray}
where $F = S_\infty+\kappa(\delta)T_\infty$. The first term is the
isotropic incoherent contribution $NI_1$ of $N$ independent atoms.
The second term enhances the previous incoherent term and also
provides the CBS cone (second term in Eq.~\eqref{TN2}). Finally,
the third term in Eq.~\eqref{Etot:terms} contributes to the
coherent forward emission as the sum of the single and double
scattering contributions.

\begin{figure}
\centering\includegraphics[width=0.95\linewidth]{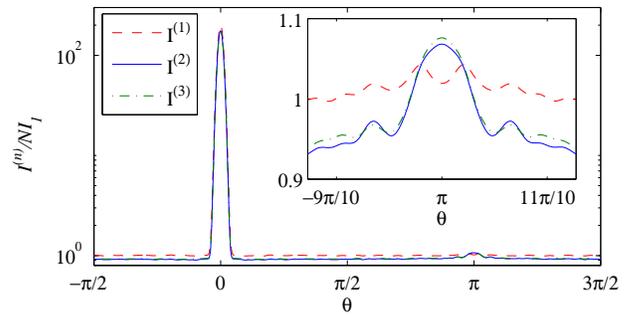}
\caption{(Color online) Far-field scattered intensity vs. $\theta$
up to the $1^{st}$, $2^{nd}$ and  $3^{rd}$ scattering order, i.e.,
$I^{(n)}=(\varepsilon_0c/2)|\sum_{j=1}^n E^{(j)}|^2$, in units of
$NI_1$. The inset is a zoom of the backscattering region (linear
scale). Simulations realized for a Gaussian spherical cloud of
on-resonant optical thickness $b_0=1$, $N=200$ particles, scaled
size $\sigma=20$ and laser detuning $\delta=0.5$. The intensity
has been averaged over $1000$
realizations.}\label{f:interference}\vspace{-0.4cm}
\end{figure}

The CBS cone reveals itself upon averaging over the pair
double scattering term~\eqref{TN2}. We first average over the
direction of the vector $\mathbf{r}_j-\mathbf{r}_m$, assuming an
atomic distribution with infinite boundaries, as for instance the
Gaussian one which is easy to parametrize. Moreover, there is
no need to know the details about the density of the cloud, since
we deal  with angular variables only. We note that although the procedure does not correspond to a rigorous
configuration average, it allows for analytical results and compares well with numerical results obtained by
configuration averages (see Fig.~\ref{f:DiagEm3o}).

This first averaging results in (see Appendix \ref{Appendix:ave}
for details):
\begin{eqnarray}\label{T}
    \langle|T_N|^2_{\mathrm{pair}}\rangle&=&\frac{1}{N^2}\sum_{j}\sum_{m\neq j}\frac{1}{k^2r_{jm}^2}\nonumber\\
    &\times &\left\{1+
    \frac{\sin[2kr_{jm}\cos(\theta/2)]}{2kr_{jm}\cos(\theta/2)}
    \right\},
\end{eqnarray}
where $\theta$ refers to the angle of $\mathbf{k}$ with respect to
the direction of $\mathbf{k}_0$. The average over the pair
distance $r_{jm}=|\mathbf{r}_j-\mathbf{r}_m|$ is the next step,
and the resulting backscattering enhancement depends on the atomic
distribution. In the next section, we discuss the CBS for Gaussian
spheres.

\subsection*{Gaussian sphere density profile}

\begin{figure}[!ht]
    \includegraphics[width=6cm]{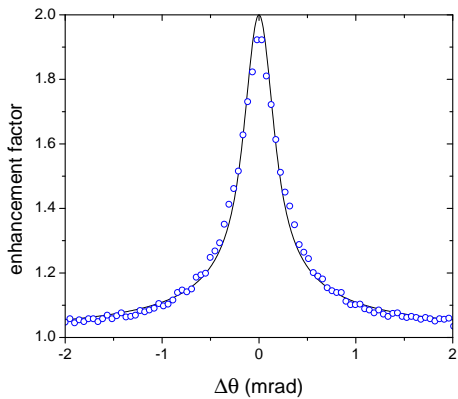}
    \caption{Experimental and theoretical CBS enhancement $E(\theta)$ for a Gaussian sphere of normalized standard
    deviation $\sigma=8098$ and $\delta=0$. The circles correspond to the experimental values reported in Fig.2 of
    Ref.~\cite{Bidel2002}, while the plain curve reproduces  Eq.~\eqref{Itot:fin}. It must be noted that the only free parameter
    is a $3\%$ adjustment of the background. }\label{fig5}\vspace{-0.4cm}
\end{figure}
As discussed previously, our multiple scattering approach is valid
for arbitrary geometries, including inhomogeneous media. Let us
illustrate this on a Gaussian sphere of standard deviation
$\sigma_R$, for which the contribution of double scattering to CBS
enhancement reads (see Appendix \ref{Appendix:gauss}):
\begin{eqnarray}
    \langle|T_N(\theta)|^2_{\mathrm{pair}}\rangle&=&\frac{1}{2\sigma^2}
    \left\{1+\frac{\sqrt{\pi}}{2}\frac{\mathrm{erf}[2\sigma\cos(\theta/2)]}{2\sigma\cos(\theta/2)}\right\}\nonumber
   \\ &=&\frac{E(\theta)}{2\sigma^2}\label{Tfin}
\end{eqnarray}
with $\sigma=k\sigma_R$. $E(\theta)$ has a maximum enhancement of
$2$ (see Fig.~\ref{fig5}) and an angular FWHM of
$\Delta\theta=2\sqrt{3}/\sigma\approx 0.55(\lambda/\sigma_R)$.

For a Gaussian sphere, the single-scattering forward contribution
gives, (see Ref.~\cite{Courteille2010}, Eq.~\eqref{SInfty}),
\begin{equation} \label{SGauss}
S_\infty(\theta)=\exp[-2\sigma^2\sin^2(\theta/2)],
\end{equation}
while the second-order forward contribution, in the limit of large
spheres $\sigma\gg 1$, is:
\begin{equation}\label{TGauss}
    T_\infty(\theta)\approx \frac{N}{4\sigma^2}\exp[-4\sigma^2\sin^2(\theta/4)].
\end{equation}
The exact expression and its derivation are given in Appendix
\ref{Appendix:Tcoher}. The total scattered intensity for a
Gaussian sphere, up to the second scattering order (see Eq.\eqref{Etot:terms}), reads:
\begin{eqnarray}\label{Itot:fin}
    I_{\mathrm{sca}}(\mathbf{r})&=& I_1 N\left\{1+\frac{b(\delta)}{4}E(\theta)+ N|F(\theta)|^2\right\},
\end{eqnarray}
where the forward contribution is given by:
\begin{eqnarray}\label{Icoher}
    F(\theta)&=& e^{-2\sigma^2\sin^2(\theta/2)}-(1+2i\delta)\frac{b(\delta)}{8}e^{-4\sigma^2\sin^2(\theta/4)},\nonumber\\
\end{eqnarray}
and $b(\delta)=b_0/(1+4\delta^2)$ with $b_0=2N/\sigma^2$ being the
resonant optical thickness for scalar light. Eq.~\eqref{Icoher}
highlights the fact that the multiple scattering expansion is
performed in orders of $b(\delta)=\sqrt{2\pi}(\sigma_R/\ell)$,
i.e. in orders of inverse scattering mean free path
$\ell=1/[\rho_0\sigma_{sc}(\delta)]$, where
$\sigma_{sc}(\delta)=4\pi/[k^2(1+4\delta^2)]$ is the scattering
cross section.

The background second-order scattering is observed to interfere
destructively with the background first-order scattering in
Fig.\ref{f:interference}, leading to an overall reduction of the
background radiation. This effect is already present in the
mathematical expression of the forward contribution given by
Eq.(\ref{Icoher}), which is expected to be the dominant term
except for the backward direction.

{\em Coherent Back-Scattering.---} Our analysis is in excellent agreement with the
experimental results of Bidel and coworkers~\cite{Bidel2002}, see
Fig.~\ref{fig5}. These authors probed the CBS cone for a large
cloud ($\sigma=8098$) of scalar optical thickness $b_0=1.93$ at
resonance, and measured an angular width of the cone of $0.50\pm
0.04$ mrad. This result is in accord with the theoretical value of
$0.46$ mrad derived from Eq.\eqref{Tfin}.

{\em Coherent Forward Scattering.---} The single-scattering forward lobe is given by the first
term in \eqref{Icoher}, and reflects the diffraction from the sample. Surprisingly, we also observe a forward
lobe for the double scattering contribution, given by the second term in \eqref{Icoher}.
The ratio between the peak intensity of the double scattering compared to that of single scattering is always given
by $b_0^2/64(4\delta^2+1)$, and the ratio of their power by $b_0^2/32(4\delta^2+1)$ for $\sigma\gg1$, independently of spatial density.

We however note that increasing the system size at constant $b_0$ and $\delta$ will increase the peak amplitude and power for
both first and second scattering orders (last $N$ factor in \eqref{Itot:fin}), yet their ratio remain constant.
This coherent forward scattering lobe could be compared to that of Refs.~\cite{Karpiuk2012,vanTiggelen1990}.
In both these works, the forward lobe appears only in the high spatial density limit close to the Anderson localization threshold,
whereas in our case, the lobe is also present in the low-density limit.

\section{Conclusion}

We have proposed an iterative multiple scattering approach, where
the radiation field at each scattering order is obtained from the
field at the atomic positions calculated at the previous order.
Provided all the eigenvalues of the iterative scattering operator
have below-unity eigenvalues, it provides a converging solution
for the multiple scattering problem. In the opposite case, the
picture of waves being scattered at one atom after the other
collapses, and the multiple scattering series becomes divergent. A
limitation of the approach is that the derivation of the $n$th
scattering order involves $n-1$ integrals over the cloud
distribution, which practically limits the efficiency of the
method to the first scattering orders for non-trivial geometries.

On the other hand, the series permits us to link observable
phenomena to particular scattering orders thus deepening our
understanding of their physical origin. As an example, for
arbitrary distributions we calculate the double-scattering
contributions to backward coherent radiation, the so-called CBS
cone.

Finally, the multiple scattering approach presented in this paper
may find applications in other many-body scattering problems. One
such example is the elucidation of the relationship between
Bragg scattering and the phenomenon of photonic bandgaps, the
first one occurring in the single and the second one in the
multiple scattering regime.


\section{Acknowledgements}

We acknowledge financial support from Research Executive Agency
(Program COSCALI, Grant No. PIRSES-GA-2010-268717), from
USP/COFECUB (project Uc Ph 123/11) and from GDRI "NANOMAGNETISM,
SPIN ELECTRONICS, QUANTUM OPTICS AND QUANTUM TECHNOLOGIES". M.T.R.
is supported by an Averro\`es exchange program. R.B. and Ph.W.C.
acknowledge the support from the Funda\c{c}\~ao de Amparo \`{a} 
Pesquisa do Estado de S\~ao Paulo (FAPESP).

\appendix

\section{Derivation of Eq.(\ref{Evetto})\label{Appendix:field}}

The radiation field can be obtained from Maxwell equations in the
presence of a polarization $\mathbf{P}$. The equations for the
Fourier component at the frequency $\omega=ck$ are:\vspace{-0.4cm}
\begin{eqnarray}
  \nabla\times \mathbf{E} &=& i\omega\mathbf{B} \label{eqM1}\\
  \nabla\times \mathbf{B} &=& -i\omega\mu_0(\epsilon_0 \mathbf{E}+\mathbf{P}) \label{eqM2}\\
   \nabla\cdot \mathbf{E} &=& -(1/\epsilon_0)\nabla\cdot \mathbf{P}\label{eqM3}
\end{eqnarray}
Taking the curl of Eq.(\ref{eqM1}) and using Eq.(\ref{eqM2}),
\begin{eqnarray}
  \nabla\times\nabla\times \mathbf{E} &=&  (\omega^2/c^2)
  [\mathbf{E}+(1/\epsilon_0)\mathbf{P}]\label{eqM12}
\end{eqnarray}
where $c=(\varepsilon_0\mu_0)^{-1/2}$ is the vacuum speed of
light. Using the identity $\nabla\times\nabla\times
\mathbf{E}=\nabla(\nabla\cdot \mathbf{E})-\nabla^2 \mathbf{E}$ and
Eq.(\ref{eqM3}), we obtain
\begin{eqnarray}
  \left(\nabla^2 +k^2\right)\mathbf{E}&=&-\frac{k^2}{\epsilon_0}\left[
  \mathbf{P}+\frac{1}{k^2}\nabla(\nabla\cdot \mathbf{P})\right].
  \label{eqM5}
\end{eqnarray}
The solution of eq.(\ref{eqM5}) is easily obtained using the
scalar Green's function $G(r)=\exp(ikr)/(ikr)$ as
\begin{equation}\label{Er}
\mathbf{E}(\mathbf{r})=i\frac{k^3}{4\pi\epsilon_0}\int
d\mathbf{r}' G(|\mathbf{r}-\mathbf{r}'|)
\left[\mathbf{P}(\mathbf{r}')+\frac{1}{k^2}\nabla(\nabla\cdot
\mathbf{P}(\mathbf{r}'))\right]
\end{equation}
or, for each components,
\begin{eqnarray}\label{Er21}
E_\alpha(\mathbf{r})&=&i\frac{k^3}{4\pi\epsilon_0}\sum_{\beta}\int
d\mathbf{r}' G(|\mathbf{r}-\mathbf{r}'|)\nonumber\\
&\times
&\left[\delta_{\alpha,\beta}+\frac{1}{k^2}\frac{\partial^2}{\partial
x'_\alpha\partial x'_\beta}\right]P_\beta(\mathbf{r}').
\end{eqnarray}
By integrating by parts and using Eq.(\ref{GvGs}), we obtain
\begin{equation}\label{Er2}
E_\alpha(\mathbf{r})=i\frac{k^3}{6\pi\epsilon_0}\sum_{\beta} \int
d\mathbf{r}' G_{\alpha,\beta} (\mathbf{r}-\mathbf{r}')
P_{\beta}(\mathbf{r}').
\end{equation}
Taking a discrete distribution of electric dipoles, with
$\mathbf{P}(\mathbf{r})=-d\sum_{j=1}^N\mathbf{b}_j\delta(\mathbf{r}-\mathbf{r}_j)$,
we obtain:
\begin{equation}\label{Ebeta}
\mathbf{E}(\mathbf{r})=-i\frac{dk^3}{6\pi\epsilon_0}\sum_{m=1}^N
\mathbf{G}(\mathbf{r}-\mathbf{r}_{m})\cdot \mathbf{b}_m.
\end{equation}

\section{Average over random angular variables for the CBS contribution\label{Appendix:ave}}

Assuming in the double scattering contribution of Eq.(\ref{TN2})
$\mathbf{k}_0=k(0,0,1)$,
$\mathbf{k}=k(\sin\theta\cos\phi,\sin\theta\sin\phi,\cos\theta)$
and
$\mathbf{r}_{jm}=r_{jm}(\sin\theta_{jm}\cos\phi_{jm},\sin\theta_{jm}\sin\phi_{jm},\cos\theta_{jm})$,
where $\mathbf{r}_{jm}=\mathbf{r}_j-\mathbf{r}_m$,  then
\begin{eqnarray}\label{phase}
    (\mathbf{k}+\mathbf{k}_0)\cdot\mathbf{r}_{jm}&=& kr_{jm}[
    \sin\theta\sin\theta_{jm}\cos(\phi_{jm}-\phi)\nonumber
    \\ &&+(1+\cos\theta)\cos\theta_{jm}].
\end{eqnarray}
By averaging over $\theta_{jm}$ and $\phi_{jm}$, we obtain
\begin{widetext}
\begin{eqnarray}\label{Tave}
    \langle|T_N|_{\mathrm{pair}}^2\rangle&=&\frac{1}{N^2}\sum_{j,m\neq
    j}\frac{1}{k^2r_{jm}^2}\nonumber\\
    &\times &
    \left\{1+
    \frac{1}{4\pi}\int_0^{2\pi}d\phi_{jm}\int_0^\pi
    d\theta_{jm}\sin\theta_{jm}
    \cos\{
    kr_{jm}\left[
    \sin\theta\sin\theta_{jm}\cos(\phi_{jm}-\phi)+(1+\cos\theta)\cos\theta_{jm}\right]\}
    \right\}.
\end{eqnarray}
Using the expression $\int_0^{2\pi}d\phi
\cos[a+b\cos(\phi-\phi')]=2\pi\cos(a)J_0(b)$, the integration over
$\phi_{jm}$ gives
\begin{equation}\label{Tave2}
    \langle|T_N|^2_{\mathrm{pair}}\rangle=\frac{1}{N^2}\sum_{j}\sum_{m\neq j}\frac{1}{kr_{jm}^2}\left\{1+
    \frac{1}{2}\int_{0}^\pi d\theta_{jm}\sin\theta_{jm}
    \cos[kr_{jm}(1+\cos\theta)\cos\theta_{jm}]J_0\left[kr_{jm}\sin\theta\sin\theta_{jm}\right]
    \right\}.
\end{equation}
Using the expression
\begin{equation}\label{Int:b}
\int_{0}^\pi d\theta\sin\theta
    \cos(a\cos\theta)J_0\left(b\sin\theta\right)=2\frac{\sin\sqrt{a^2+b^2}}{\sqrt{a^2+b^2}},
\end{equation}
we obtain
\begin{equation}\label{T:app}
    \langle|T_N|_{\mathrm{pair}}^2\rangle=\frac{1}{N^2}\sum_{j}\sum_{m\neq j}\frac{1}{kr_{jm}^2}\left\{1+
    \mathrm{sinc}[2kr_{jm}\cos(\theta/2)].
    \right\}
\end{equation}
where $\mathrm{sinc}(z)=\sin(z)/z$.
\end{widetext}

\section{Derivation of Eq.(\ref{Tfin}) \label{Appendix:gauss}}
Let's consider the integral
\begin{equation}\label{Int}
    I=\int d\mathbf{r}_1\rho(\mathbf{r}_1)\int
    d\mathbf{r}_2\rho(\mathbf{r}_2)
    f(|\mathbf{r}_1-\mathbf{r}_2|).
\end{equation}
Changing integration variables from $\mathbf{r}_1$ and
$\mathbf{r}_2$ to $\mathbf{R}=(\mathbf{r}_1+\mathbf{r}_2)/2$ and
$\mathbf{s}=\mathbf{r}_1-\mathbf{r}_2$,
\begin{equation}\label{Intb}
    I=\int d\mathbf{R}\int
    d\mathbf{s}\rho(\mathbf{R}-\mathbf{s}/2)\rho(\mathbf{R}+\mathbf{s}/2)
    f(|\mathbf{s}|).
\end{equation}
Assuming a Gaussian distribution,
$\rho(\mathbf{r})=\rho_0\exp(-r^2/2\sigma_R^2)$, since
$|\mathbf{R}+\mathbf{s}/2|^2+|\mathbf{R}-\mathbf{s}/2|^2=2R^2+s^2/2$,
in polar coordinates the integral (\ref{Int}) becomes
\begin{eqnarray}\label{Int2}
    I&=&\frac{2N^2}{\pi\sigma_R^6}\int_0^\infty dR
    R^2e^{-R^2/\sigma_R^2}\int_0^\infty ds s^2 e^{-s^2/4\sigma_R^2}
    f(s)\nonumber\\
    &=&\frac{4N^2}{\sqrt{\pi}}\int_0^\infty dx x^2 e^{-x^2}
    f(2\sigma_R x).
\end{eqnarray}
Taking
\begin{equation}
f(s)=\frac{1}{k^2s^2}\left\{1+\frac{\sin[2ks\cos(\theta/2)]}{2sk\cos(\theta/2)}\right\},
\end{equation}
the integral is
\begin{eqnarray}\label{Int3}
    I&=&\frac{N^2}{\sigma^2\sqrt{\pi}}\int_0^\infty dx e^{-x^2}
    \left[1+\frac{\sin(ax)}{ax}\right]\nonumber\\
    &=&\frac{N^2}{2\sigma^2}
    \left[
    1+\frac{\sqrt{\pi}}{a}\mathrm{erf}(a/2)
    \right]
\end{eqnarray}
where $\sigma=k\sigma_R$  and $a=4\sigma\cos(\theta/2)$.

\section{Derivation of Eq.(\ref{TGauss}) \label{Appendix:Tcoher}}

Let us consider the coherent contribution
$T_\infty(\mathbf{k},\mathbf{k}_0)$ of Eq.(\ref{Tcont}) and
introduce $\mathbf{R}=(\mathbf{r}_1+\mathbf{r}_2)/2$ and
$\mathbf{s}=\mathbf{r}_1-\mathbf{r}_2$, so that

\begin{eqnarray}
T_\infty(\mathbf{k},\mathbf{k}_0)&=&\frac{1}{ikN}\int
d\mathbf{R}\int
d\mathbf{s}\rho(\mathbf{R}+\mathbf{s}/2)\rho(\mathbf{R}-\mathbf{s}/2)\nonumber
\\ &&\times
s^{-1}e^{i(\mathbf{k}_0-\mathbf{k})\cdot
\mathbf{R}+i(\mathbf{k}_0+\mathbf{k})\cdot
\mathbf{s}/2+iks}\label{Tcoh2}
\end{eqnarray}
For a Gaussian distribution the double integral factorizes,
\begin{eqnarray}
T_\infty(\mathbf{k},\mathbf{k}_0)&=&\frac{\rho_0^2}{ikN}\int
d\mathbf{R}e^{-R^2/\sigma_R^2+i(\mathbf{k}_0-\mathbf{k})\cdot
\mathbf{R}}\label{Tcoh3}
\\ &&\times\int d\mathbf{s}s^{-1}
e^{-s^2/4\sigma_R^2+i(\mathbf{k}_0+\mathbf{k})\cdot
\mathbf{s}/2+iks}.\nonumber
\end{eqnarray}
Assuming
$\mathbf{k}=(\sin\theta\cos\phi,\sin\theta\sin\phi,\cos\theta)$
and $\mathbf{k}_0=k(0,0,1)$, the first integral yields:
\begin{eqnarray}
I_1&=&\rho_0\int
d\mathbf{R}e^{-R^2/\sigma_R^2+i(\mathbf{k}_0-\mathbf{k})\cdot
\mathbf{R}}=\frac{N}{2\sqrt{2}}e^{-\sigma^2(1-\cos\theta)/2}\nonumber
\end{eqnarray}
where $\sigma=k\sigma_R$. The second integral of Eq.(\ref{Tcoh3}),
after integration over the angular variables, yields:
\begin{eqnarray}
I_2&=&\frac{2N}{i\sqrt{2\pi}\sigma^2 \cos(\theta/2)}\int_0^\infty
dx e^{-x^2/4+i\sigma x}\sin[\sigma\cos(\theta/2)x]\nonumber
\end{eqnarray}
Using the expression
\begin{eqnarray}
\int_0^\infty dx
e^{-x^2/4+iax}&&\sin(bx)=\frac{\sqrt{\pi}}{2}\Big\{
-e^{4ab}\left[\mathrm{erfi}(a-b)-i\right]\nonumber
\\ &&+\mathrm{erfi}(a+b)-i\Big\}e^{-(a+b)^2}
\end{eqnarray}
we obtain, from the above equations
\begin{eqnarray}\label{Tcoh4}
T_\infty(\theta)&=&\frac{Ne^{-2\sigma^2(1+\cos\theta/2)}}{4i\sigma^2
\cos\theta/2}
\Bigg\{\mathrm{erfi}\left[\sigma(1+\cos\theta/2)\right]
\\ &&-e^{4\sigma^2\cos\theta/2}\mathrm{erfi}\left[\sigma(1-\cos\theta/2)\right]+i\left(e^{4\sigma^2\cos\theta/2}-1\right)
\Bigg\}\nonumber
\end{eqnarray}
Notice that $T_\infty(\pi)=N\exp(-2\sigma^2)$. For large $\sigma$
and near the forward direction, we get:
\begin{eqnarray}\label{Tcoh5}
T_\infty(\theta)&\approx &\frac{N}{4\sigma^2}
e^{-4\sigma^2\sin^2(\theta/4)}\approx\frac{b_0}{8}
e^{-4\sigma^2\sin^2(\theta/4)}.
\end{eqnarray}


\end{document}